# A Search for Satellites around Ceres

A. Bieryla[1,2], J. Wm. Parker[2], E.F. Young[2], L. A. McFadden[3], C. T. Russell[4], S. A. Stern[2], M. V. Sykes[5] and B. Gladman[7]

[1] Harvard-Smithsonian Center for Astrophysics, 60 Garden Street, MS-6, Cambridge, MA 02138, USA abieryla@cfa.harvard.edu
[2] Southwest Research Institute, 1050 Walnut Street, Suite 300, Boulder, CO 80302, USA
[3] Goddard Spaceflight Center, Greenbelt, MD 20771, USA
[4] Institute of Geophysics and Planetary Physics, University of California, Los Angeles, CA 90095, USA
[5] Planetary Science Institute, 1700 East Fort Lowell, Suite 106, Tucson, AZ, 85719 USA
[6] Department of Physics and Astronomy, University of British Columbia, 6224 Agricultural Road, Vancouver, BC V6T1Z1, Canada

## Abstract

We conducted a satellite search around the dwarf planet 1 Ceres using *Hubble Space Telescope* and ground-based Palomar data. No candidate objects were found orbiting Ceres in its entire stability region down to ~500km from the surface of Ceres. Assuming a satellite would have the same albedo as Ceres, which has a visual geometric albedo of 0.07—0.10, our detection limit is sensitive to satellites larger than 1-2 km in diameter.

## 1. Introduction

1 Ceres is the smallest dwarf planet and the first known and largest main-belt asteroid. Previous observations by *Hubble Space Telescope (HST)* reveal Ceres to be an oblate spheroid with axes a=487 km and b=455km, with a surface topography that is apparently relaxed (Thomas et al. 2005). The shape and rotation suggests Ceres is differentiated with an ice-rich mantle (Thomas et al. 2005).

Observations of satellites around minor planets are valuable because they can be used to probe the physical properties of the components. The total mass of the system can be determined from the relative orbit of the satellite, and the individual masses can be determined if the orbits of both components around the barycenter can be measured. Further, if the sizes of one or both components can be measured (either directly or from thermal observations), then density and albedo can be determined. Such data can give clues to the composition and interior structure properties of a body, which can be considered in context with its taxonomy group as determined from spectroscopy, if available. Asteroids with satellites also allow for the study of natural collisions that are relevant to the formation and evolution of asteroids (Merline et al. 2002). As of January 2011, there were approximately 193 small bodies with known satellites. This includes 37 near-Earth asteroids, 13 Mars crossing asteroids, 73 main-belt asteroids, 4 Trojan asteroids, and 66 trans-Neptunian objects (Johnston 2011). To date, no results from any dedicated Ceres satellite searches have been published.

The existence, or lack thereof, of satellites around Ceres is also of great interest to NASA's Dawn mission, which launched on September 27, 2007 and will arrive at Ceres in February 2015 after first rendezvousing with Vesta in 2011. The Dawn spacecraft will orbit Ceres for several months at altitudes of 700 to 5900 km (Russell et al. 2006), which is well inside the satellite stability region. In addition to providing additional targets for physical studies by the Dawn instruments, satellites could affect the orbit design in order to accommodate observations and spacecraft safety. Therefore, advance knowledge of the environment around Ceres is critical to mission planning.

In this paper we report on our project to search for satellites around Ceres using the *Hubble Space Telescope* (*HST*) and the Hale 5-meter telescope at Palomar Observatory. This combination of space- and ground-based datasets allowed us to search the full extent of the satellite stability region around Ceres and particularly to high resolution in the inner region.

## 2. Observations and Analysis

We made observations of Ceres using the High Resolution Channel (HRC) of the Advanced Camera for Surveys (ACS) on *HST* to study the physical properties of Ceres (Parker et al. 2004, Thomas et al. 2005, Li et al. 2006) and to search for satellites. We observed Ceres with the *HST* ACS-HRC in December 2003 and January 2004. At that time, Ceres was 1.65 AU from the Earth, producing a scale of 30 km/pixel, and an angular diameter of ~ 0.8 arcsec (about 31 pixels across). A total of 267 individual images of Ceres were obtained during nine *HST* orbits. The first six orbits were consecutive, covering the extent of the 9.1 hr rotation period, and the remaining three orbits were obtained at different times in the following month to improve rotational phase coverage and spatial resolution. Most of the exposures on the six consecutive orbits used a 12×12 arcsec (512×512 pixel) subarray of the HRC to provide shorter readout time and in order to fit more images into the *HST* memory. At the end of four of these six orbits, "long *V*" (F555W filter) exposures were taken using the full 25×25 arcsec field of view of the ACS-HRC in order to search for possible satellites of Ceres. These four observations were made with a 15-second exposure time and a CR-SPLIT=3 to robustly remove cosmic rays. A satellite in these images would have a typical orbital speed of less than ~0.25 km/s (we later show that we could detect faint satellites as close as 0.4 arcsec from Ceres), so it will move significantly less than a resolution element over the course of a single 15-second CR-SPLIT exposure. Thus, we can combine the three sub-exposures for each observation to remove cosmic rays without accidentally removing any potential satellites (the observations tracked at Ceres' apparent motion). The images were processed using the standard *HST* pipeline data reduction and IRAF and IDL programs.

The *HST* data were able to obtain high resolution close to Ceres, but the field of view was considerably smaller than Ceres' Hill sphere. As discussed by Hamilton and Burns (1991), the Hill sphere has a radius $R_H = (\mu/3)^{1/3}R$, where $R$ is the circular heliocentric orbit of the primary body, and $\mu = m_p/(m_\odot + m_p)$ is the reduced mass-parameter. For Ceres, $R_H = 2.2 \times 10^5$ km which translates to an apparent Hill angular radius of about 189 arcsec during the epoch of our *HST* observations. Initially-circular prograde orbits remain bound out to about half the Hill region, $R_H/2 = 1.1 \times 10^5$ km or 95 arcsec, while initially-circular retrograde orbits are stable out to almost the full Hill sphere. The *HST* images cover an area which is only the inner ~2.5% for prograde orbits and 0.6% for retrograde orbits of the stability region.

To get complete coverage of the satellite stability region around Ceres, we also obtained ground-based observations on 2006 July 27 with the Large Format Camera on the Hale 5m telescope at Palomar Observatory. Each CCD of the camera covers an area of roughly 720×360 arcsec (with the larger dimension in the E-W direction). The images were binned 2x2, producing a plate scale of 0.36 arcsec/pixel. At the time, Ceres was 2.02 AU from the Earth, resulting in a physical scale of 530 km/pixel, and angular radii of the prograde stability region and Hill sphere of 74 arcsec and 149 arcsec, respectively. Five one-minute exposures using the SDSS *r'* filter were taken in succession over a period of 9 minutes. The motion of Ceres was small but noticeable throughout, covering about 4 arcsec between the first and last exposures. Ceres moved 0.5 arcsec during a single exposure, which was far less than the seeing disk of FWHM~2.2 arcsec, so the motion of any satellite would be readily detectable by eye over the series of images with little trailing loss in each image.

We performed similar search methods and analyses on both the *HST* and the Palomar images sets. We searched for objects co-moving with Ceres by blinking the normally processed images with various levels of contrast, and then by blinking the same set of images after performing median-filter subtraction (high-pass filter or unsharp masking) to remove large-scale variations due to scattered light from Ceres. These filtered images allowed us to search closer to Ceres and to a fainter detection limit. In the *HST* images, there were no other objects (e.g., stars) in the images with Ceres, so the search involved simply looking for any object that was persistent among the images taken at different times. In the larger Palomar images, many stars were present, and the images were manually blinked to look for any objects that moved at the same rate as Ceres relative to the stars.

We determined the detection limits of our search by the common method of planting "fake" objects in the images. These objects were modeled to have the same shape as the image point spread function (PSF), and were scaled to a range of magnitudes and placed in the images at random positions with motion consistent with Ceres. The images were then re-examined and any detected objects were noted. In this way, we were able to produce detection efficiency curves as a function of magnitude and distance from Ceres for each instrument. This also provided a second pass through the data to look for real objects that would show up as detections that weren't in the list of fake objects.

# 3. Results

We did not find any satellites around Ceres. Our results for our *HST* detection efficiency are shown in Figure 1 and our Palomar detection efficiency in Figure 2. The conversion to diameter annotated along the top axis assumes a satellite with the same albedo as Ceres, i.e., we used the observed magnitude and diameter of Ceres to normalize the diameter scale along the top axis.

Our *HST* detection limits are valid to as close as ~0.4 arcsec (a projected distance of ~480 km) off the surface of Ceres. For objects in circular orbits about Ceres, 480 km is well inside the Roche limit of Ceres. Any satellite with an orbit smaller than that could have escaped our detection due to scattered light from Ceres. However, we also examined the shorter-exposure F555W images that were taken at different epochs; although those images did not go as deep (a factor of 5 less exposure time) and covered ¼ of the area as the 15-second images, they provided some additional check for satellites that could have been obscured by scattered light from Ceres in the 15 second images. Also, any satellite in a low inclination orbit could have been obscured by occultations/transits with Ceres, but we calculate that based on the timing of our observations relative to the range of orbital periods, any satellite with an orbit larger than ~825 km would have appeared in at least two of our images. There is a 35% chance a satellite on a smaller orbit (surface < R < 825 km) would not have appeared in at least two of the 15-second images, but those closer distances were obscured by scattered light anyway in the 15-second exposures.

With the Palomar data, we are not able to detect objects as close to Ceres due to lower resolution and Ceres being over-exposed, but those ground-based images were able to cover an area well beyond the Hill radius, providing a complementary dataset to the *HST* observations covering the full stability region of potential satellites.

We are able to say (at the 90% detection level) that there are no objects larger than ~1 km in diameter orbiting Ceres in the *HST* images from ~500 km above the surface out to a distance of ~15,000 km; that size limit corresponds to a magnitude of ~21.5. With the same detection level of 90%, we can say that there are no objects larger than ~2 km in diameter (magnitude ~21) orbiting Ceres in the entire stability region. The two datasets together imply that Ceres does not have any satellites larger than about 1—2 km in diameter. These results assume a satellite would have the same albedo as Ceres; if it has an albedo different than that of Ceres, then the diameter limit of our search would be correspondingly different by a factor of 1/sqrt(albedo), e.g., if the albedo of a satellite were 4 times larger than the albedo of Ceres, then our detection size limit would be 2 times smaller. Note that our results are not dependent on the specific numeric values for the albedos of the satellite or Ceres, but only depend on the *relative* values of their albedos. For reference, published values for the visual geometric albedo of Ceres are in the range 0.07-0.10 (Li et al. 2006; Millis et al. 1987; Tedesco 1989).

The result that Ceres does not have any satellites larger than ~1 km is interesting. Unlike Dawn's first target Vesta, which has a large asteroid family associated with it, probably arising from the large impact resulting in the crater covering Vesta's southern hemisphere, Ceres has no associated asteroid family and exhibits no large impact structures on its surface. Ceres is alone among the asteroids in having a hydrostatic equilibrium shape, which is maintained by its gravity and ice-rich composition. This may offer some insight into why Ceres has no associated family or satellites: ejecta would likely be ice-rich and because Ceres' orbit is entirely within the snow-line around 3 AU, they may not survive against solar radiation, much like a disintegrating cometary body. Also, craters on an ice-rich surface would likely flatten out over time because there would be a quicker relaxation time on a surface that is mechanically weak.

*Acknowledgements*. This work was supported by the HST grant from STScI (HST-GO-09748). We would like to thank Jake Hanson for his help on the occultation analysis.

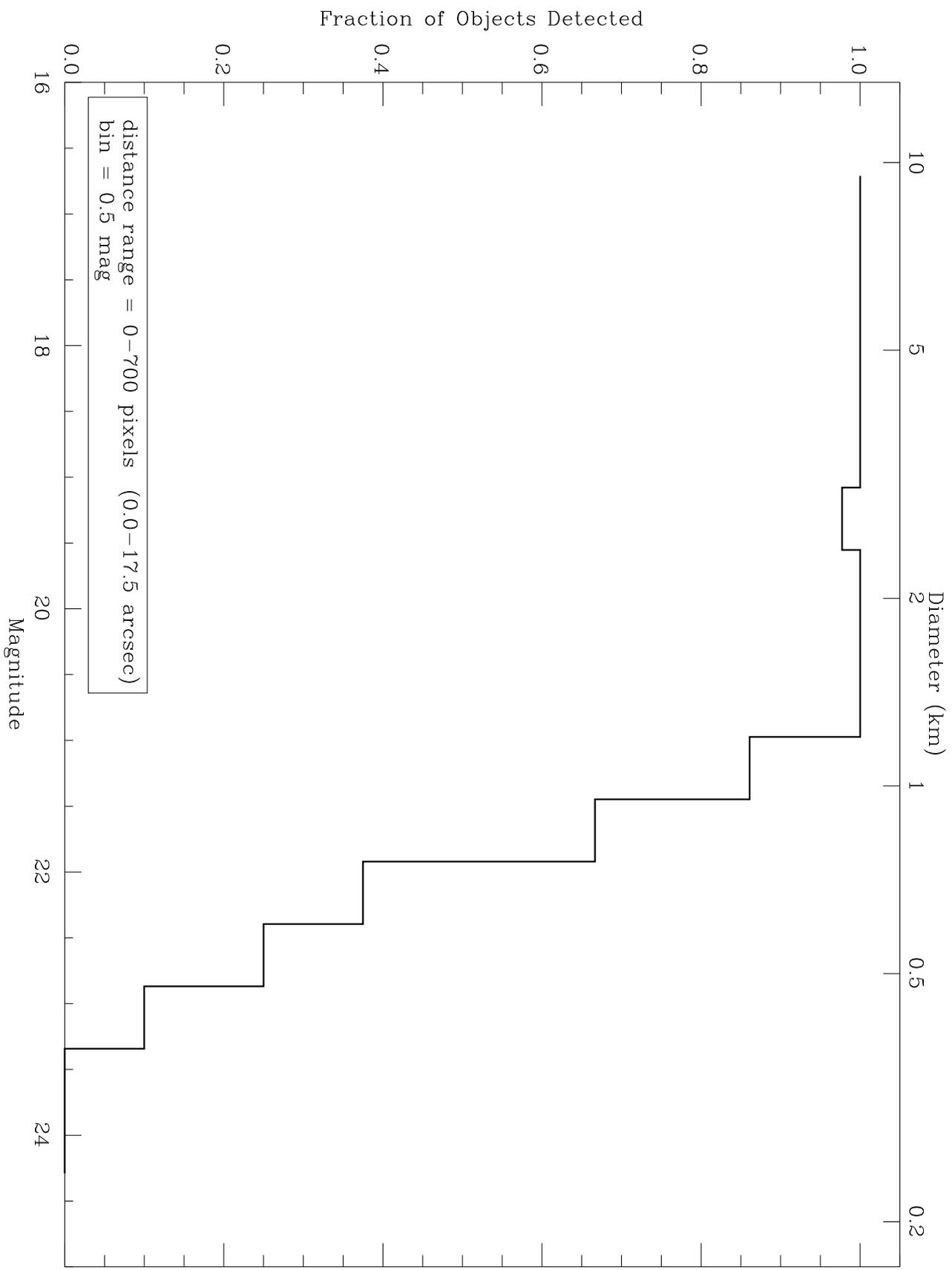

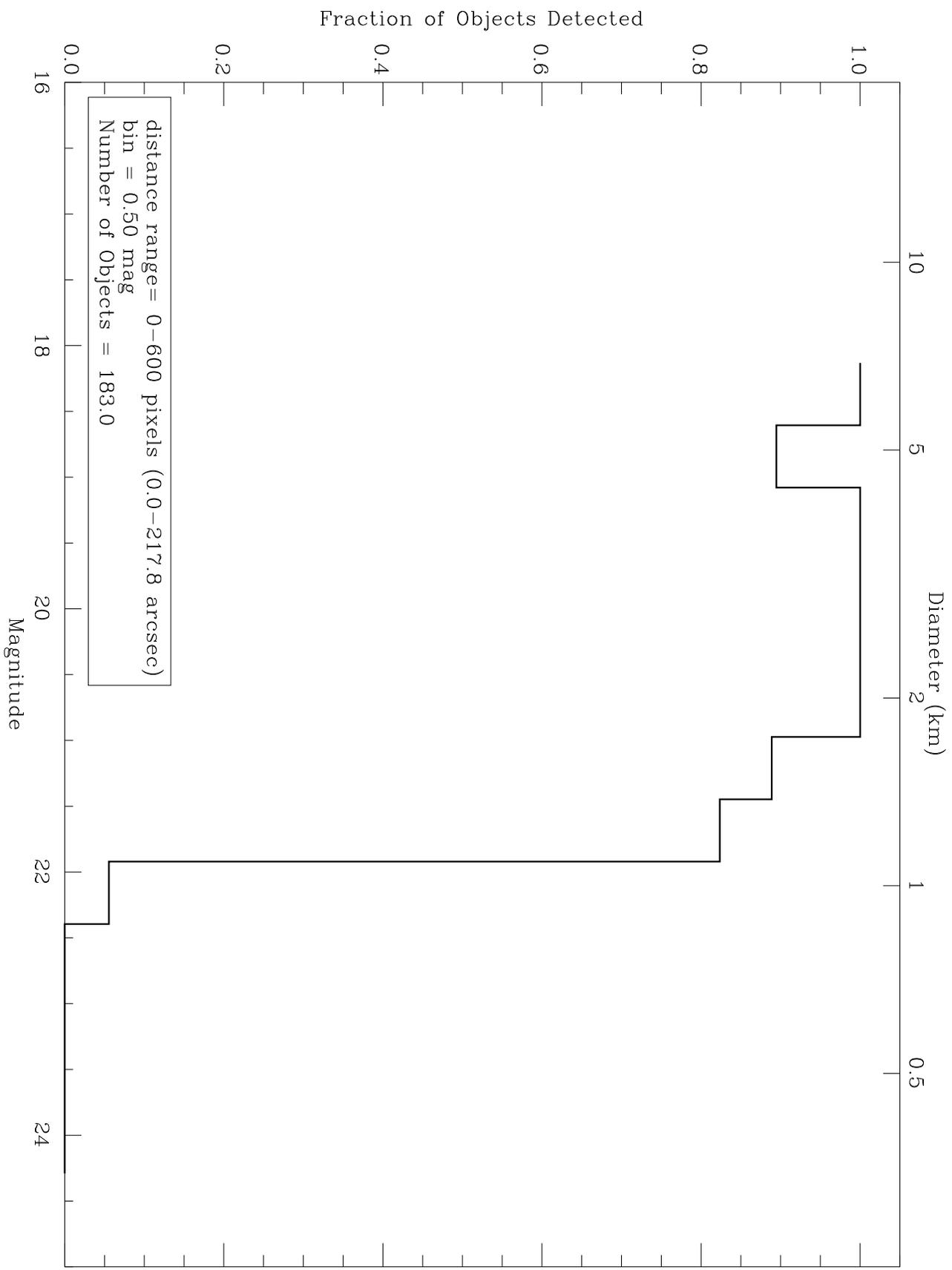